\documentclass[%
 aip,
 amsmath,amssymb,
 reprint,%
]{revtex4-1}

\usepackage{graphicx}% Include figure files
\usepackage{dcolumn}% Align table columns on decimal point
\usepackage{bm}% bold math
\usepackage[utf8]{inputenc}
\usepackage[T1]{fontenc}
\usepackage{etoolbox}
\usepackage[dvipsnames]{xcolor}
\usepackage[super]{nth}

\makeatletter
\def\@email#1#2{%
 \endgroup
 \patchcmd{\titleblock@produce}
  {\frontmatter@RRAPformat}
  {\frontmatter@RRAPformat{\produce@RRAP{*#1\href{mailto:#2}{#2}}}\frontmatter@RRAPformat}
  {}{}
}%
\makeatother
\begin{document}

\preprint{AIP/123-QED}

\title[Critical slowing of the spin and charge density wave order in thin film Cr following photoexcitation]{Critical slowing of the spin and charge density wave order in thin film Cr following photoexcitation}
\author{Sheena K.K. Patel}
 \email{skp008@ucsd.edu, efullerton@ucsd.edu}
 \affiliation{Center for Memory and Recording Research, University of California San Diego, 9500 Gilman Dr., La Jolla, CA 92093}
 \affiliation{Department of Physics, University of California San Diego, 9500 Gilman Dr., La Jolla, CA 92093}
 
\author{Oleg Yu. Gorobtsov}%
\affiliation{Department of Materials Science and Engineering, Cornell University, Ithaca, NY 14853}

\author{Devin Cela}%
 \affiliation{Department of Physics, University of California San Diego, 9500 Gilman Dr., La Jolla, CA 92093}

\author{Stjepan B. Hrkac}
 \affiliation{Department of Physics, University of California San Diego, 9500 Gilman Dr., La Jolla, CA 92093}

\author{Nelson Hua}
 \affiliation{Department of Physics, University of California San Diego, 9500 Gilman Dr., La Jolla, CA 92093}

\author{Rajasekhar Medapalli}
 \affiliation{Center for Memory and Recording Research, University of California San Diego, 9500 Gilman Dr., La Jolla, CA 92093}

\author{Anatoly G. Shabalin}
 \affiliation{Department of Physics, University of California San Diego, 9500 Gilman Dr., La Jolla, CA 92093}

\author{James Wingert}
 \affiliation{Department of Physics, University of California San Diego, 9500 Gilman Dr., La Jolla, CA 92093}

\author{James M. Glownia}
 \affiliation{Linac Coherent Light Source, SLAC National Accelerator Laboratory, Menlo Park, California 94025}

\author{Diling Zhu}
 \affiliation{Linac Coherent Light Source, SLAC National Accelerator Laboratory, Menlo Park, California 94025}

\author{Matthieu Chollet}
 \affiliation{Linac Coherent Light Source, SLAC National Accelerator Laboratory, Menlo Park, California 94025}

\author{Oleg G. Shpyrko}
 \affiliation{Department of Physics, University of California San Diego, 9500 Gilman Dr., La Jolla, CA 92093}

\author{Andrej Singer}
 \affiliation{Department of Materials Science and Engineering, Cornell University, Ithaca, NY 14853}

\author{Eric E. Fullerton}
 \affiliation{Center for Memory and Recording Research, University of California San Diego, 9500 Gilman Dr., La Jolla, CA 92093}

\date{29 February 2024}% It is always \today, today,

\begin{abstract}
We report on the evolution of the charge density wave (CDW) and spin density wave (SDW) order of a chromium film following photoexcitation with an ultrafast optical laser pulse. The CDW is measured by ultrafast time-resolved x-ray diffraction of the CDW satellite that tracks the suppression and recovery of the CDW following photoexcitation. We find that as the temperature of the film approaches a discontinuous phase transition in the CDW and SDW order, the time scales of recovery increase exponentially from the expected thermal time scales. We extend a Landau model for SDW systems to account for this critical slowing with the appropriate boundary conditions imposed by the geometry of the thin film system. This model allows us to assess the energy barrier between available CDW/SDW states with different spatial periodicities. 
\end{abstract}

\maketitle

%\begin{quotation}
%The ``lead paragraph'' is encapsulated with the \LaTeX\ 
%\verb+quotation+ environment and is formatted as a single paragraph before the first section heading. 
%(The \verb+quotation+ environment reverts to its usual meaning after the first sectioning command.) 
%Note that numbered references are allowed in the lead paragraph.
%
%The lead paragraph will only be found in an article being prepared for the journal \textit{Chaos}.
%\end{quotation}

\section{\label{sec:Intro}Introduction}

Significant condensed matter research is focused on correlated electron materials in which emergent phenomena arise from the interplay between degrees of freedom such as charge, spin, and lattice. \cite{tokura2017emergent,morosan2012strongly,basov2017towards} Ultrafast pump and probe techniques have opened possibilities for understanding how these degrees of freedom first uncouple and then recouple and evolve dynamically, quantifying the energy scales in the system and advancing the possibility of manipulating such phenomena for technological application. Ultrafast dynamics and transient nonequilibrium states from optical excitation have been observed in magnetic materials, \cite{kirilyuk2010ultrafast,hellman2017interface,kimel2019writing,wang2020ultrafast} including Ni, \cite{beaurepaire1996ultrafast,cheng2023terahertz} Py, \cite{panda2023ultrafast} rare earth transition metal alloys and heterostructures, \cite{radu2011transient,mangin2014engineered} FePt, \cite{reid2018beyond,lambert2014all} and Sr$_2$Cu$_3$O$_4$Cl$_2$. \cite{shan2024dynamic} On longer time scales, the degrees of freedom recouple and reach equilibrium at the initial temperature as the sample cools. During the latter process, studies in some magnetic materials, such as for La$_{1/3}$Sr$_{2/3}$FeO$_3$ \cite{zhu2018unconventional} and MnWO$_4$, \cite{niermann2015critical} have shown a critical slowing down of the recovery near discontinuous phase transitions.

\begin{figure*}
    \includegraphics[width=\textwidth]{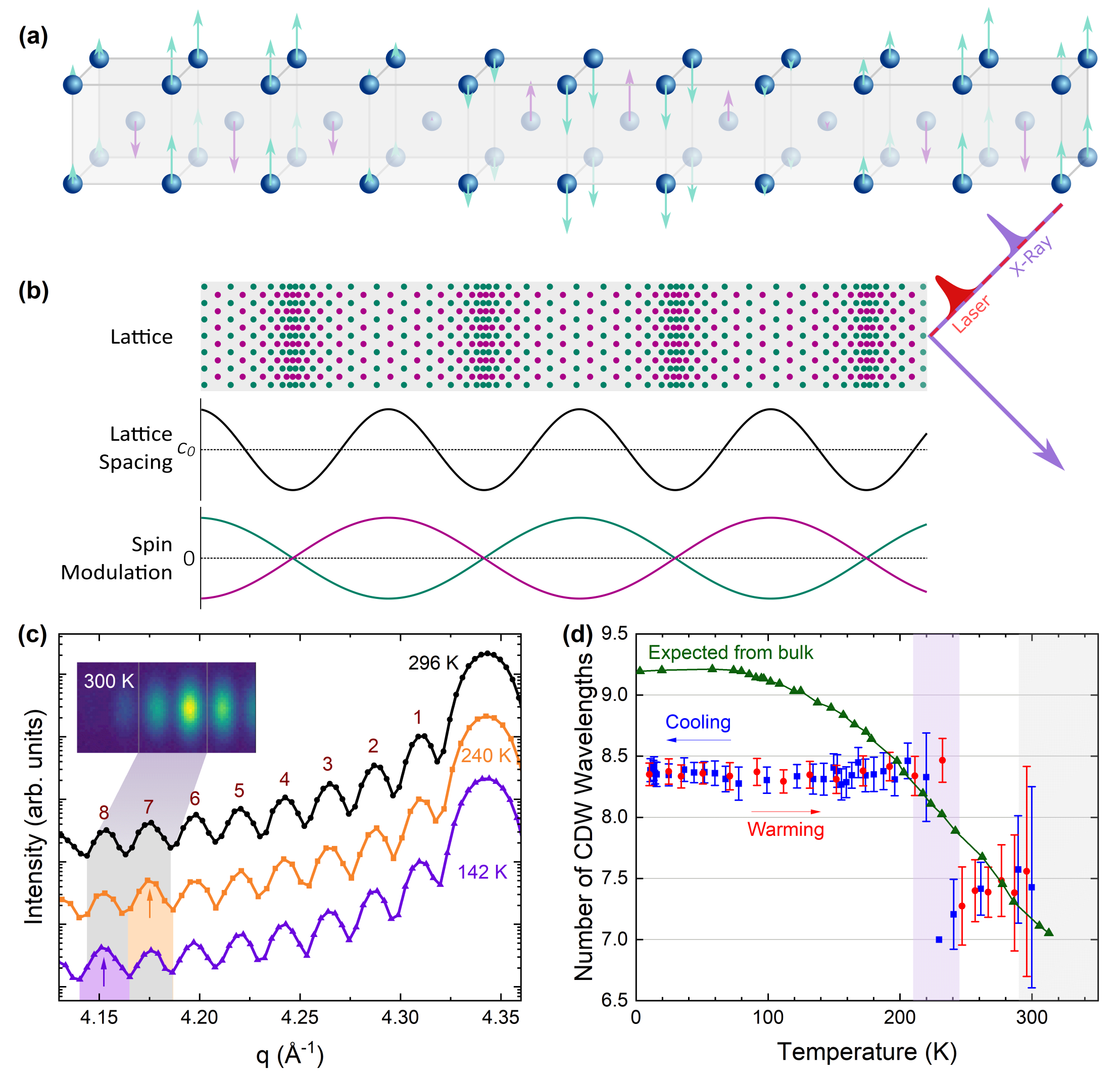}
    \caption{(a) Representation of the crystal structure and transverse SDW of Cr. The wavelength of the spin modulation is not to scale. (b) Exaggerated schematic representation of the SDW and CDW ordering in Cr. Green and pink represent vertex and body-centered atoms in the lattice, respectively. The x-ray scattering geometry required to measure the CDW satellite peak is depicted. (c) X-ray diffraction $\theta - 2\theta$ scans of the 002~Bragg peak and Laue oscillations at 296~K (black circles), 240~K (orange squares), and 100~K (purple triangles). The Laue fringes are labeled, and arrows indicate the added intensity where the CDW satellite peaks appear on fringe~7 at 240~K and fringe~8 at 142~K. The inset shows an averaged 2D detector image from the diffraction experiment at LCLS with scattering geometry aligned to the \nth{7} fringe at 300~K. Adjacent fringes are visible in the detector image due to large mosaic spread in the film. (d) The expected and observed number of CDW wavelengths in the 28-nm Cr film. The expected value (green triangles) is calculated from the bulk SDW wave vector assuming a 28 nm film (data from Ref.~\onlinecite{werner1967temperature}). The observed value for the film during cooling (blue squares) and warming (red circles) was found in previous work. \cite{singer2016phase} The purple shaded section indicates the hysteretic spin and charge reorientation region from 210~K to 245~K. The gray shaded section indicates the region above the Néel temperature.}
    \label{fig:Fig1}
\end{figure*}

This work demonstrates this critical slowing of the magnetic and structural ordering in elemental chromium (Cr) thin films. In bulk, Cr is an antiferromagnet that exhibits an incommensurate spin density wave (SDW) and charge density wave (CDW) that leads to a periodic lattice distortion as a second harmonic of its SDW (see Figs.~\ref{fig:Fig1}(a) and \ref{fig:Fig1}(b)). \cite{fawcett1988spin,zabel1999magnetism,hu2022real} The wavelength of the SDW/CDW evolves continuously with temperature during cooling. Previous studies have shown that the CDW order in thin film Cr takes on discrete wavelengths due to the boundary conditions set by the substrate and surface interfaces. \cite{kummamuru2008electrical,singer2015condensation,singer2016phase} Exploiting these discrete allowable wavelengths and the time resolution of the free electron laser x-ray source, we resolved the dynamics on ultrafast time scales following photoexcitation. \cite{singer2016photoinduced,wingert2020direct,gorobtsov2021femtosecond,li2022phonon} We observed ultrafast SDW demagnetization and the transformation of the static CDW into a dynamic coherent phonon with transient enhancement of the CDW order at low laser fluences. \cite{singer2016photoinduced,gorobtsov2021femtosecond}

Here, we study the recovery of SDW and CDW orders in thin film Cr following ultrafast photoexcitation at different initial temperatures. We observe two distinct time scales to the CDW recovery. One time scale is consistent with thermal recovery for a metallic film. The second time scale ranges from thermal recovery time scales to two orders of magnitude longer as the ground state temperature of the system approaches the phase transition in which the CDW and SDW undergo a discrete change in wavelength. This critical slowing of the recovery in the time domain provides a method of assessing the energy scales of different states in the system by comparing the experimental results to a Landau model for the Cr CDW/SDW systems to thin films with fixed boundaries.

\section{\label{sec:ExpMeth}Experimental Methods}

\subsection{\label{subsec:Sample}Sample preparation and characteristics}

The sample studied is a 28-nm Cr~[001] film grown on MgO~(001) as detailed in Refs.~\onlinecite{singer2016phase,singer2016photoinduced,wingert2020direct,gorobtsov2021femtosecond,li2022phonon}. The Néel transition is broad and centered at $T_N = 290 \pm 5$~K, as determined by transport measurements. The SDW and CDW wave vectors are directed perpendicular to the film surface and pinned such that the CDW exhibits a half-integer number of wavelengths spanning the film from substrate to surface. \cite{zabel1999magnetism,singer2015condensation,singer2016phase,soh2011spintronics,logan2012antiferromagnetic,fullerton1996spin,fullerton2003hysteretic} The number of nodes depends on the film thickness and the temperature. The temperature-induced transformation between different spatial periodicities is discontinuous and displays hysteresis with temperature. \cite{kummamuru2008electrical,singer2015condensation,singer2016phase,fullerton2003hysteretic,jaramillo2007microscopic} The CDW can be observed by x-ray diffraction as satellite peaks around the 002~Bragg peak.

Figure~\ref{fig:Fig1}(c) shows x-ray diffraction data of the 002~Bragg peak and Laue oscillations measured at three temperatures, 296~K, 240~K, and 142~K, while cooling. The CDW satellite peaks are present and aligned on the \nth{8} or \nth{7} Laue fringes, indicating 8.5 or 7.5~CDW wavelengths fit into the film at 142~K or 240~K, respectively. \cite{singer2015condensation,singer2016phase} At 296~K, the film is expected to be near or above the Néel temperature and there is little satellite peak intensity. \cite{singer2016phase} In Fig.~\ref{fig:Fig1}(d), we present the expected number of CDW wavelengths in a 28 nm film using the bulk SDW wave vector if the pinning at the interfaces is ignored (green triangles), given the temperature dependence of the wave vector in Cr from Ref.~\onlinecite{werner1967temperature}. We also show the previously determined number of CDW wavelengths in this film between the film interfaces from x-ray diffraction data from Ref.~\onlinecite{singer2016phase}. There are 7.5 or 8.5~wavelengths of the CDW observed in this film, with a hysteretic change between these two states between 210 and 245~K, indicating a mixed phase with domains exhibiting one of the two states. \cite{singer2016phase}. Given the boundary pinning and quantization of the CDW wavelength, the observed values align well with the expectation from bulk down to about 150~K. Below this temperature, the bulk wave vector plateaus such that the expected number of wavelengths in the film is almost 9.25~CDW wavelengths but the film does not exhibit a transition to a state with 9.5~CDW wavelengths. Near room temperature, the expectation from bulk raises the possibility of a transition to 6.5~CDW wavelengths (with a satellite peak expected on the \nth{6} Laue fringe) near $T_N$, which, as described below, is supported in this study.

\subsection{\label{subsec:trXRD}Time-resolved x-ray diffraction}

Time-resolved x-ray diffraction measurements were conducted at the XPP instrument at the Linac Coherent Light Source (LCLS) x-ray free-electron laser, with 8.9~keV x-ray pulses of duration 15~fs. The sample was excited by 800~nm optical pulses with a duration of 45~fs and fluences of 4.2~mJ/cm$^2$, 7.4~mJ/cm$^2$, and 9.2~mJ/cm$^2$ at initial sample temperatures ranging from 130~K to 300~K. The inset in Fig.~\ref{fig:Fig1}(c) shows a detector image where the experimental geometry is aligned to fringe~7, and we can observe the neighboring peaks with reduced intensity captured by the Ewald sphere. The time-dependent x-ray diffraction signal of several of the Laue fringes and interfering CDW satellite peaks was measured to a maximum time delay of 400~ns following photoexcitation to determine the evolution of the charge ordering in the film.

\section{\label{sec:Results}Results}

\begin{figure*}
    \includegraphics[width=\textwidth]{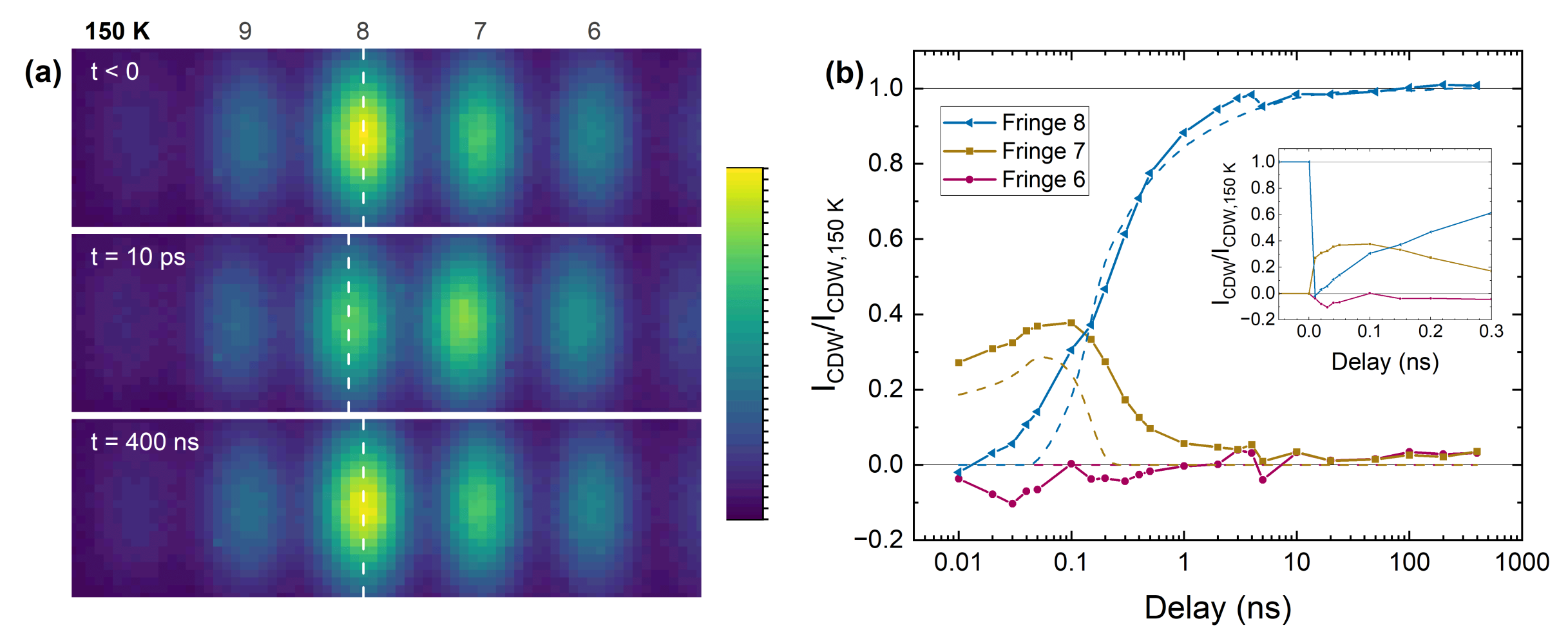}
    \caption{(a) Averaged 2D detector images of the diffraction experiment aligned to the \nth{8} fringe at 150~K before the laser pulse (top), at 10~ps after the laser pulse (middle), and at 400~ns after the laser pulse (bottom). The fringe locations shift due to thermal expansion, and the intensities change due to the change in CDW ordering. (b) CDW intensity on fringe~7 (gold squares) and fringe~8 (blue triangles) following photoexcitation with fluence 7.4~mJ/cm$^2$ normalized to the initial amplitude on fringe 8. Little change in amplitude is seen in the fringe~6 amplitude (pink circles). Dashed lines indicate the expected CDW amplitude assuming a quasi-static evolution of CDW ordering with the cooling of the lattice. Inset: CDW amplitudes on the three fringes on a linear scale in time.}
    \label{fig:Fig2}
\end{figure*}

We will focus our discussion on the response at 10 ps and longer after photoexcitation. The sub-10-ps response is discussed in detail in Refs.~\onlinecite{singer2016photoinduced,wingert2020direct,gorobtsov2021femtosecond,li2022phonon}. Following photoexcitation, a lattice expansion is observed through a shift in the Bragg peak and Laue fringes consistent with lattice heating through electron-phonon coupling, as shown in Fig.~\ref{fig:Fig2}(a). This shift allows the temperature of the sample to be monitored. In addition, the amplitudes of some of the fringes also change, depending on the initial sample temperature and the laser fluence. This change in amplitude is attributed to a change in the CDW ordering and the appearance, shift, or disappearance of the CDW satellite peak. An example is shown in Fig.~\ref{fig:Fig2}(b), which shows the time-resolved CDW satellite peak intensity on the \nth{6}, \nth{7}, and \nth{8} Laue fringes following photoexcitation with laser fluence 7.4~mJ/cm$^2$ at an initial sample temperature of 150 K. This is plotted on a log scale in time, with the inset showing the initial response on a linear scale. The CDW intensities are normalized to the initial CDW intensity on the \nth{8} fringe. In the ground state at 150~K, the CDW has 8.5~wavelengths between the film interfaces, and the CDW satellite peak intensity appears on the \nth{8} Laue fringe. Within 10~ps, the lattice temperature increases by about 125~K to 275~K \cite{singer2016photoinduced,li2022phonon} (see supplementary material Fig.~S1), which heats the system to where the CDW satellite peak intensity has completely disappeared from the \nth{8} fringe. The intensity of the \nth{8} fringe increases and recovers back to its initial configuration within about 1~ns, a time scale that is consistent with the thermal recovery of a metallic film.

Figure~\ref{fig:Fig2}(b) shows that within 10 ps, the satellite peak appears with a smaller intensity on the \nth{7} fringe, consistent with the preferred spin orientation at 275~K. The intensity of the \nth{7} fringe initially increases before decreasing back to its initial intensity. This can be explained by the change in intensity of the CDW satellite peak as a function of temperature, which decreases to zero at $T_N$. \cite{singer2016phase} At a lattice temperature of 275~K, the CDW peak would be expected to appear entirely on fringe~7, and as the system cools, the CDW peak intensity will increase on this fringe before switching to fringe~8 as the sample cools past the hysteretic region. We also show that there is little change in intensity of fringe 6. The dashed lines in Fig.~\ref{fig:Fig2}(b) show the expected CDW intensities on each fringe calculated quasi-statically from the film temperature (determined by the Bragg peak shift) and the ground state CDW peak locations and amplitude. \cite{singer2015condensation,singer2016phase} The time scales and intensities follow these expectations reasonably well.  

\begin{figure}
    \includegraphics[width=\columnwidth]{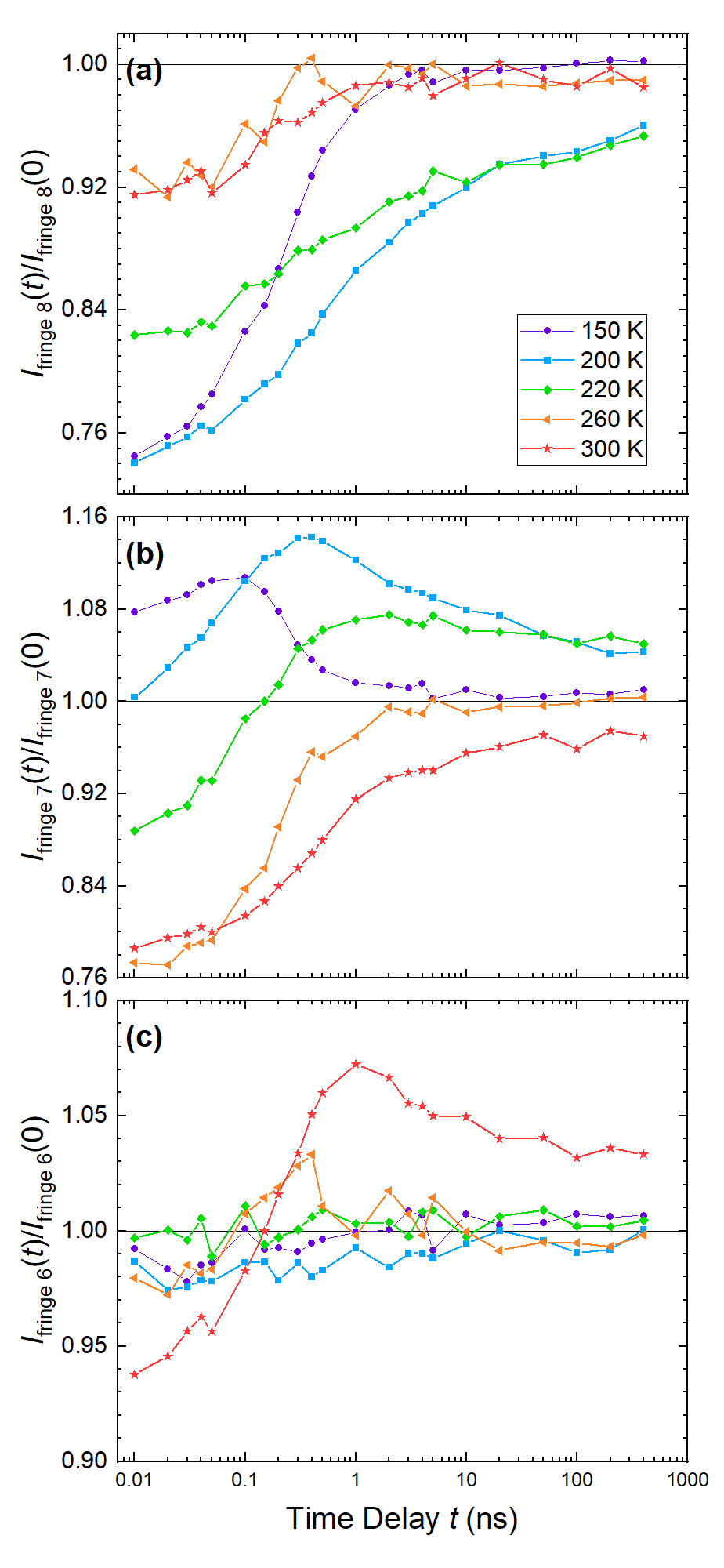}
    \caption{Normalized intensities of (a) fringe~8, (b) fringe~7, and (c) fringe~6 following photoexcitation with fluence 7.4~mJ/cm$^2$ from 150~K (purple circles), 200~K (blue squares), 220~K (green diamonds), 260~K (orange triangles), and 300~K (red stars). Fringe intensities are normalized to their initial intensity prior to photoexcitation.}
    \label{fig:Fig3}
\end{figure}

Figure~\ref{fig:Fig3}(a) shows the evolution of the intensity of the \nth{8} Laue fringe for five initial sample temperatures from 150~K to 300~K with a laser fluence of 7.4~mJ/cm$^2$. In this case, an intensity greater than 1 indicates enhanced fringe intensity following photoexcitation, and an intensity less than 1 indicates suppressed fringe intensity. As we observed in Fig.~\ref{fig:Fig2}, the CDW satellite peak intensity at 150~K disappears from this fringe and recovers consistent with a thermal recovery model. At 200~K, the initial drop in intensity is similar to that at 150~K. Yet the recovery of the CDW satellite peak onto this fringe occurs much slower: in fact, it has not recovered back to the initial state by 400~ns, our longest measurement time. This is even more dramatic at 220~K, though the initial drop in intensity is lower due to the decreasing amplitude of the CDW peak and the initial mixed state of either 7.5 or 8.5~CDW wavelengths. At 260~K and 300~K, where the ground state is 7.5~CDW wavelengths, there is a small decrease in intensity on thermal time scales consistent with what is observed on all the measured Laue fringes at all temperatures and is likely due to a temperature gradient and strain in the film and not related to CDW order. 

The normalized \nth{7} fringe intensity for the same five temperatures and pump fluence are shown in Fig.~\ref{fig:Fig3}(b). Dynamics at 150~K are as described earlier. At 200~K, the lattice temperature has increased to above $T_N$, thus few domains have the 7.5~CDW wavelengths configuration. As the system cools, the domains initially recover back into the higher temperature phase with that 7.5~CDW wavelength configuration, so the CDW satellite peak appears on the \nth{7} fringe before disappearing as the system cools further below the transition. As we saw in Fig.~\ref{fig:Fig3}(a), at this starting temperature, the system’s recovery proceeds much slower than at 150~K, and there are domains persisting in the higher temperature phase out to at least 400~ns. At 220~K, there is initially a CDW satellite peak contribution to the \nth{7} fringe which disappears when pumped and recovers back with greater intensity and the system has not recovered back to its initial state at 400~ns. Above the hysteretic transition region, at 260 K, the CDW satellite peak, which initially appears entirely on the \nth{7} fringe, disappears with the laser pulse as the temperature reached is well above $T_N$ before recovering back consistent with thermal recovery. Interestingly, at 300~K, we see evidence of the persistence of magnetic order, with a significant initial drop in intensity. The recovery here occurs on a longer time scale and remains incomplete after 400~ns, indicating that we may be near another critical transition here that is not easily determined from static measurements.

Since we expect the CDW satellite peak to appear primarily on the \nth{7} and \nth{8} fringes for this film based on previous static x-ray diffraction measurements, \cite{singer2016phase} the intensity of the other fringes, including the \nth{6}, should have little response to photoexcitation. Figure~\ref{fig:Fig3}(c) shows this response. At the lower temperatures, this proceeds as expected. However, at higher temperatures, there is some indication of CDW satellite peak appearing on the \nth{6} fringe, which would be in agreement with Ref.~\onlinecite{wingert2020direct}. At 260~K, there is a small increase in intensity during the thermal recovery. At 300~K, when the system is expected to be excited to above $T_N$, there is a significant drop and then a rise in intensity, followed by a slower recovery. This suggests that there is another hysteretic transition region around 300 K where we would expect domains in the sample to be in either the 7.5 or 6.5~CDW wavelength configuration (Fig.~\ref{fig:Fig1}) and there is a slowing of the recovery of order that mirrors what was seen at lower temperature. This further suggests that the static Néel transition in this film is higher than 290~K or is quite broad and demonstrates that the recovery of magnetic order to the state with 6.5~CDW wavelengths begins on the order of tens of ps after photoexcitation, when the lattice temperature is above 400~K.

Similar measurements were made with different pump fluence (see supplementary material Fig.~S2-S6). These demonstrate similar time scales for recovery despite different evolution of intensity, which is a result of different total heating from the laser pulse. Thus, the critical slowing depends on the initial sample temperature but not on the strength of the photoexcitation.

\section{\label{sec:Discussion}Discussion}

\subsection{\label{subsec:Ebarrier}Energy barrier}

\begin{figure}
    \includegraphics[width=\columnwidth]{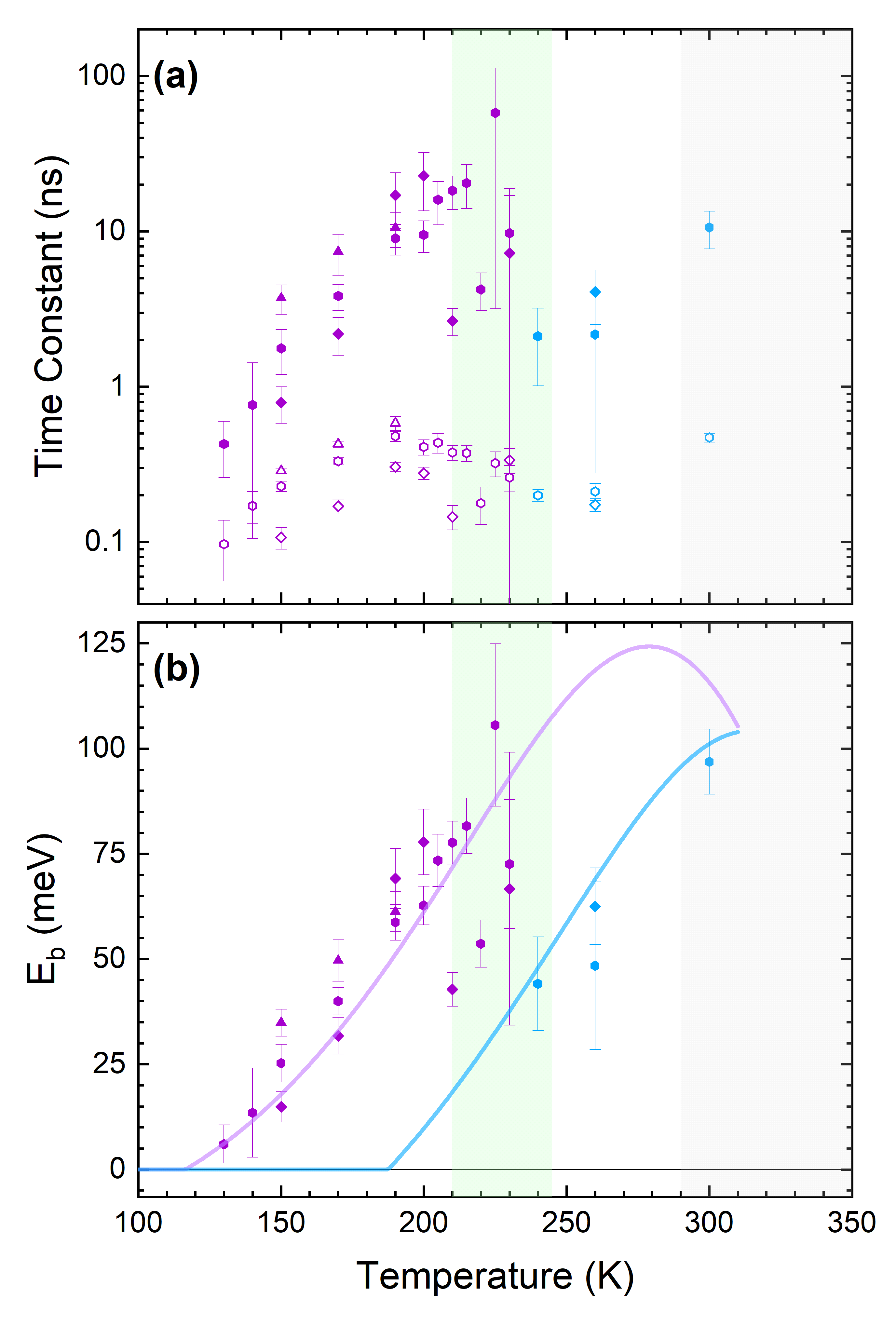}
    \caption{(a) Time constants $\tau_\text{fast}$ (open data points) and $\tau_\text{slow}$ (solid data points) for the recovery of CDW order to its initial state from a double exponential fit. Fits were made to the time-dependent CDW amplitude on fringe~8 below the transition (purple) and fringe~7 above the transition (blue). (b) Energy barrier calculated from $\tau_\text{slow}$ for the transition from $N_\text{CDW}=7.5$ to 8.5 (purple data points) or from $N_\text{CDW}=6.5$ to 7.5 CDW (blue data points) assuming an attempt time of 250~ps. The solid lines are the energy barriers calculated for both transitions from the Landau free energy model with added boundary pinning terms. Fitting constants and energy barriers are included for data at laser fluences 4.2~mJ/cm$^2$ (diamonds), 7.4~mJ/cm$^2$ (circles), and 9.2~mJ/cm$^2$ (triangles).}
    \label{fig:Fig4}
\end{figure}

We describe the recovery of structural order with two distinct time constants $\tau_\text{fast}$ and $\tau_\text{slow}$ to follow an exponential decay function: 
\begin{equation}
\Delta\text{CDW} \sim A_1 \, \mathrm{e}^{-\frac{t}{\tau_\text{fast}}} + A_2 \, \mathrm{e}^{-\frac{t}{\tau_\text{slow}}} + C \, .
\end{equation}
The constant $C$ is needed within the hysteretic region to account for the presence of a mixed phase. Some example fits are shown in supplementary material Fig.~S6 and the fitting parameters for all data sets are presented in Table~S1. The time constants from the fits for all fluences and temperatures are shown in Fig.~\ref{fig:Fig4}(a). As we pump from temperatures below and approaching the hysteretic region,  $\tau_\text{fast}$ remains in a range of 100~ps to 400~ps while $\tau_\text{slow}$ increases by about two orders of magnitude. Due to the maximum measurable delay time of 400~ns, the full recovery for some temperatures could not be measured, which negatively affects the fits in the transition region, resulting in larger uncertainty and variability for $\tau_\text{slow}$. The purple data points indicate fits to the fringe~8 intensities, giving a time constant for recovery to the 8.5~CDW wavelengths state from high temperature. The blue data points indicate fits to the fringe~7 intensities, giving a time constant for recovery to the 7.5~CDW wavelength state from high temperature. We find that $\tau_\text{fast}$ is on the order of hundreds of ps and representative of the thermal recovery process following photoexcitation. Another recovery mechanism described by $\tau_\text{slow}$ strongly depends on the initial temperature and significantly slows down approaching the transition region between different CDW wavelengths, independent of laser fluence. Just above the transition region, the recovery to the 7.5~CDW wavelength state is dominated by the thermal recovery, but $\tau_\text{slow}$ increases upon approach of a second hysteretic transition region, coincidentally near $T_N$.

\begin{figure}
    \includegraphics[width=\columnwidth]{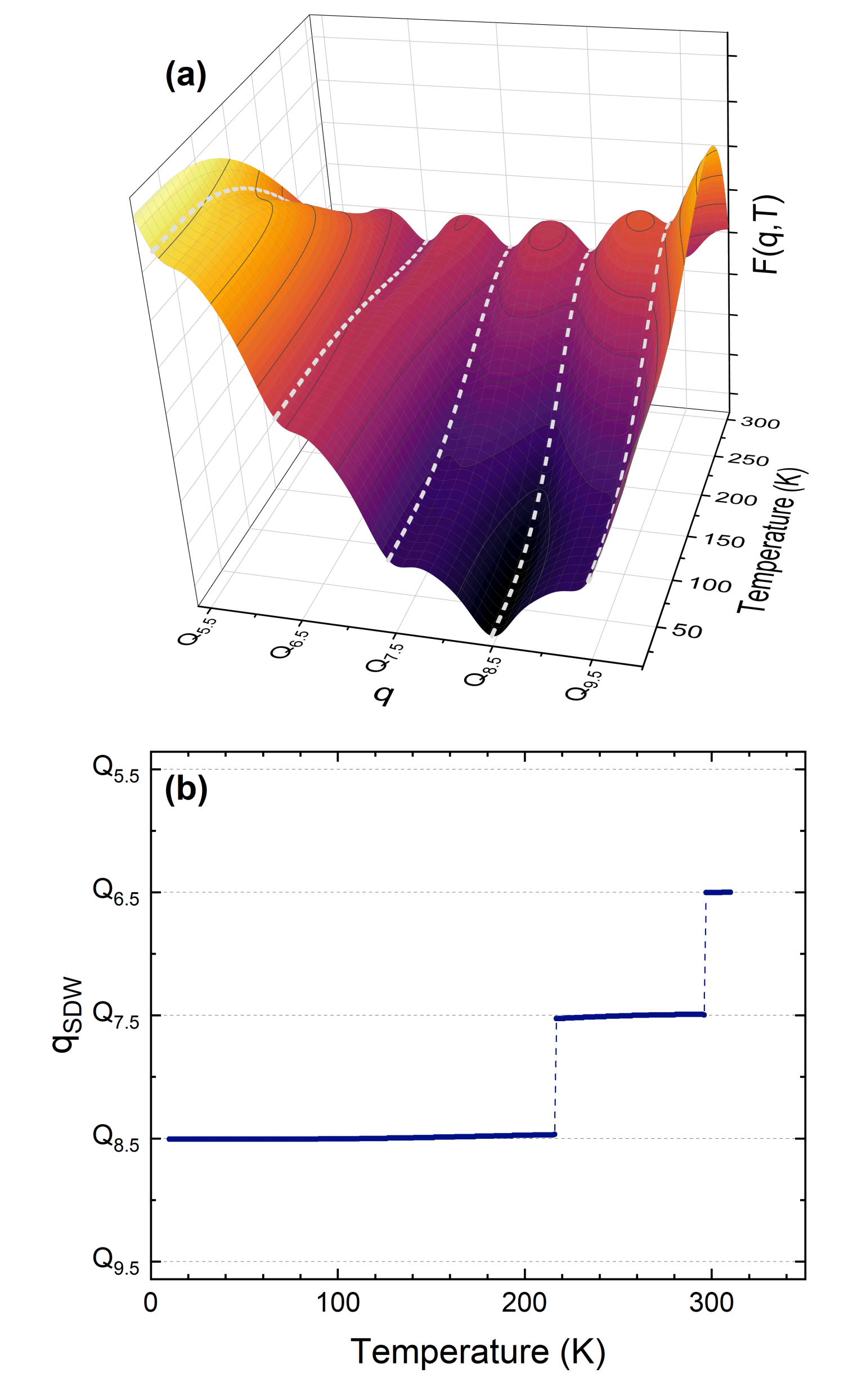}
    \caption{(a) Landau free-energy $F(q,T)$ model of the SDW/CDW wave vector. Dashed lines show the potential for the wave vectors corresponding to 9.5, 8.5, 7.5, 6.5, and 5.5~CDW wavelengths in the film. (b) The preferred ground state wave vector of the system as a function of temperature found from the minimum of the Landau free-energy surface. $Q_i$ refers to the SDW wave vector corresponding to $i$ CDW wavelengths in the film.}
    \label{fig:Fig5}
\end{figure}

The slowing down of the recovery of electronic order on approach of a weakly first-order transition has been observed previously in La$_{1/3}$Sr$_{2/3}$FeO$_3$, as a result of two possible charge orders with similar potentials. \cite{zhu2018unconventional} In thin film Cr, the pinning of the SDW/CDW at the interfaces and the discretized number of wavelengths between the interfaces gives multiple distinct spin and charge configurations of the film. The ground state at any temperature is given by configuration with the minimum potential, and the recovery time, then, reflects the energy required to reconfigure the spins throughout the depth of the film back into the appropriate lower temperature state. In an energy landscape, we can consider energy wells at each possible spin configuration, with excitation into a state which prefers fewer CDW wavelengths. As the system recovers thermally, the potential surface evolves back to its initial state, however there is an activation energy required to reconfigure each SDW/CDW domain back to its initial state. This energy barrier can be calculated using an Arrhenius model,
\begin{equation}
\ln\frac{\tau_\text{slow}}{\tau_0} = \frac{E_b}{k_B T} \, ,
\end{equation}
where $\tau_0$ is the attempt time. Figure~\ref{fig:Fig4}(b) shows the calculated energy barriers with an attempt time assumed to be $\tau_0=250$~ps for two reorientations: from 7.5~CDW wavelengths to 8.5~CDW wavelengths for purple data points and from 6.5~CDW wavelengths to 7.5~CDW wavelengths with blue data points. The energy barriers vary approximately linearly with temperature within the relevant temperature range, which will result in the exponential slowing down of the recovery of electronic order. The energy barrier appears to go to zero for the 7.5~wavelength to 8.5~wavelength transition at around 125~K, though this depends on the attempt time used for the calculation (see supplementary material Fig.~S7).

\subsection{\label{subsec:Landau}Landau model of the free energy}

The free energy density as a function of depth for a film with a SDW/CDW can be described with the Landau formalism in Refs.~\onlinecite{li2022phonon,mcmillan1977microscopic,jaramillo2009breakdown,feng2015itinerant}:
\begin{eqnarray}
F &=& a_1 \, \psi^2 + a_2 \, \psi^4 + a_3 \, \xi | \nabla \psi | ^2 \notag \\
& & + a_4 \, \psi^2 | \nabla \phi - q_0 |^2 + a_5 \, \psi^4 \cos 2\phi \, ,
\end{eqnarray}
where $\psi \mathrm{e}^{i\phi}$ is the SDW order parameter with amplitude $\psi$ and phase $\phi$, $q$ is the SDW nesting wave vector, and $a_i$ is the coefficient for the \textit{i}\textsuperscript{th} term. The first two terms follow from the Landau form for expansion of the free energy and includes the temperature dependence of the free energy with $a_{1,2} \propto (T-T_N)$. The third term adds energy with spatial variation of the order parameter, which we assume to be zero here. The fourth term accounts for the energy cost to change the SDW wave vector from the natural wave vector. The fifth term describes an energetic preference for the SDW to be commensurate with the lattice. We integrate this over the depth of the film and add a term which describes potential wells that restrict the available wave vectors to values that force a half-integer number of CDW wavelengths:
\begin{widetext}
\begin{equation}
F(q) = L \left( a_1 \, \psi^2 + a_2 \, \psi^4 + a_3 \, \xi | \nabla \psi | ^2 + a_4 \, \psi^2 | q - q_0 |^2 \right) \notag + a_5 \, \psi^4 \frac{\sin 2 L q}{2q} + a_6 \, G \left( 1-\frac{Q_j}{q} \right) ,
\end{equation}
\end{widetext}
%\begin{eqnarray}
%F(q) &=& L \left( a_1 \, \psi^2 + a_2 \, \psi^4 + a_3 \, \xi | \nabla \psi | ^2 + a_4 \, \psi^2 | q - q_0 |^2 \right) \notag \\
%& & + a_5 \, \psi^4 \frac{\sin 2 L q}{2q} + a_6 \, G \left( 1-\frac{Q_j}{q} \right) ,
%\end{eqnarray}
where $L$ is the film thickness and $G$ enforces the boundary conditions with allowed wave vectors $Q_j$ corresponding to $N_\text{CDW} =j$. The $q$-dependence of the free energy lies in the fourth, fifth, and sixth terms, which are used to model the Landau free energy surface shown in Fig.~\ref{fig:Fig5}(a), with dashed white lines at the energies of the states with 5.5, 6.5, 7.5, 8.5, and 9.5~CDW wavelengths. Figure~\ref{fig:Fig5}(b) shows the minimized SDW wave vector at the minimum energy at each temperature from this model, correctly reproducing the temperature dependence of the SDW wave vector in this film, with preferred ground state $Q_{8.5}$ below 220~K, $Q_{7.5}$ between 220~K and 300~K, and $Q_{6.5}$ above 300~K. The bulk value of $T_N=311$~K was used.

With this energy surface, we are able to explain the measured CDW response following photoexcitation. With the initial laser pulse, the preferred SDW/CDW order has changed. As the film cools past the switching temperature, there is an energy barrier to reforming the order. This energy barrier can be found from the surface in Fig.~\ref{fig:Fig5}(a), and has been plotted for the switch from $Q_{7.5}$ to $Q_{8.5}$ with a purple solid line and from $Q_{6.5}$ to $Q_{7.5}$ with a blue solid line in Fig.~\ref{fig:Fig4}(b), consistent with the data for our film. These energy barriers increase with temperature in the measured ranges, resulting in the critical slowing that we have observed.

\section{\label{sec:Conclusion}Conclusion}

In conclusion, we have observed the slowing down of the recovery of CDW order following photoexcitation from temperatures that approach a discontinuous hysteretic reorientation of spin and charge in thin film Cr to several orders of magnitude longer than expected from thermal recovery. An Arrhenius model allows us to make an estimate of the energy barrier between these states. We have extended the Landau SDW model to account for the boundary pinning observed in films and describe both the temperature dependence of the SDW and CDW ordering and the evolution of the potential wells with temperature.

\begin{acknowledgments}
The work was supported by the U.S. Department of Energy, Office of Science, Office of Basic Energy Sciences, under Contracts No. DE-SC0018237 (S.K.K.P, R.M., and E.E.F), DE- SC0001805 (D.C., S.B.H., N.H., A.G.S., J.W., A.S., and O.G.S.), and DE- SC0019414 (O.Y.G. and A.S.). Use of the Linac Coherent Light Source (LCLS), SLAC National Accelerator Laboratory, is supported by the U.S. Department of Energy, Office of Science, Office of Basic Energy Sciences under Contract No. DE-AC02-76SF00515.
\end{acknowledgments}

\section*{Data Availability Statement}

Raw data were generated at Linac Coherent Light Source (LCLS), SLAC National Accelerator Laboratory. Derived data supporting the findings of this study are available from the corresponding author upon reasonable request.

\nocite{*}
\bibliography{aipsamp}% Produces the bibliography via BibTeX.

\end{document}

\maketitle

\section{Thermal evolution due to heating from the laser}

\begin{figure*}
    \centering
    \includegraphics[width=0.7\textwidth]{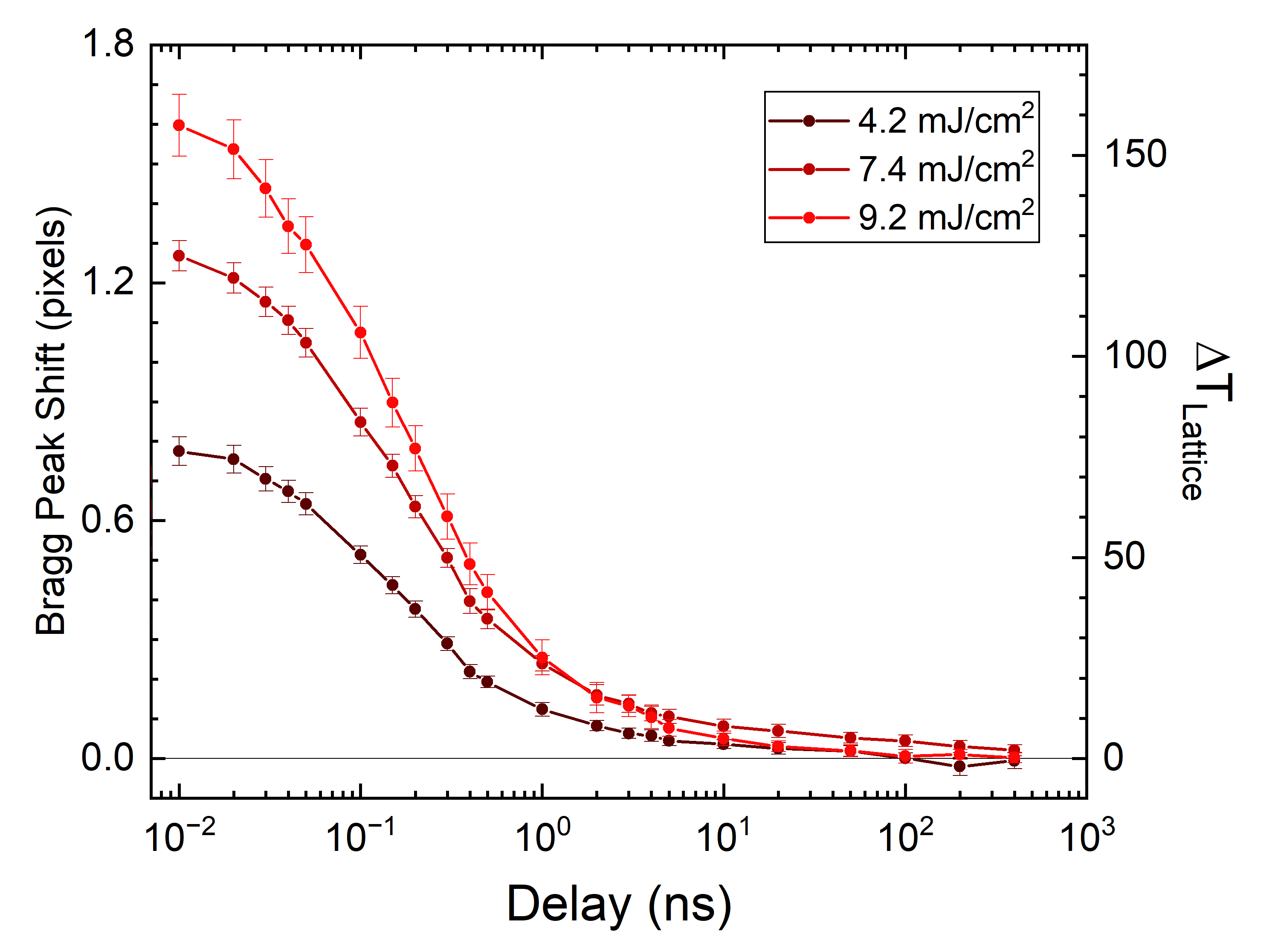}
    \caption{The shift in pixels of the Bragg peak and fringes and the associated lattice temperature following photoexcitation with three different laser fluences.}
    \label{fig:FigS1}
\end{figure*}

Figure~\ref{fig:FigS1} shows the shift of the Bragg peak and fringes on the 2D detector for each of the three laser fluences used. Data were averaged for all data sets measured. The corresponding change in temperature for the lattice was determined in previous work.\footnote{Supplementary Information for A. Singer, S. K. K. Patel, R. Kukreja, V. Uhlíř, J. Wingert, S. Festersen, D. Zhu, J. M. Glownia, H. T. Lemke, S. Nelson, M. Kozina, K. Rossnagel, M. Bauer, B. M. Murphy, O. M. Magnussen, E. E. Fullerton, and O. G. Shpyrko, “Photoinduced enhancement of the charge density wave amplitude,” Phys. Rev. Lett. {\bf 117}, 056401 (2016).}

\section{Fringe intensities at all temperatures and fluences}

\begin{figure*}
    \centering
    \includegraphics[width=\textwidth]{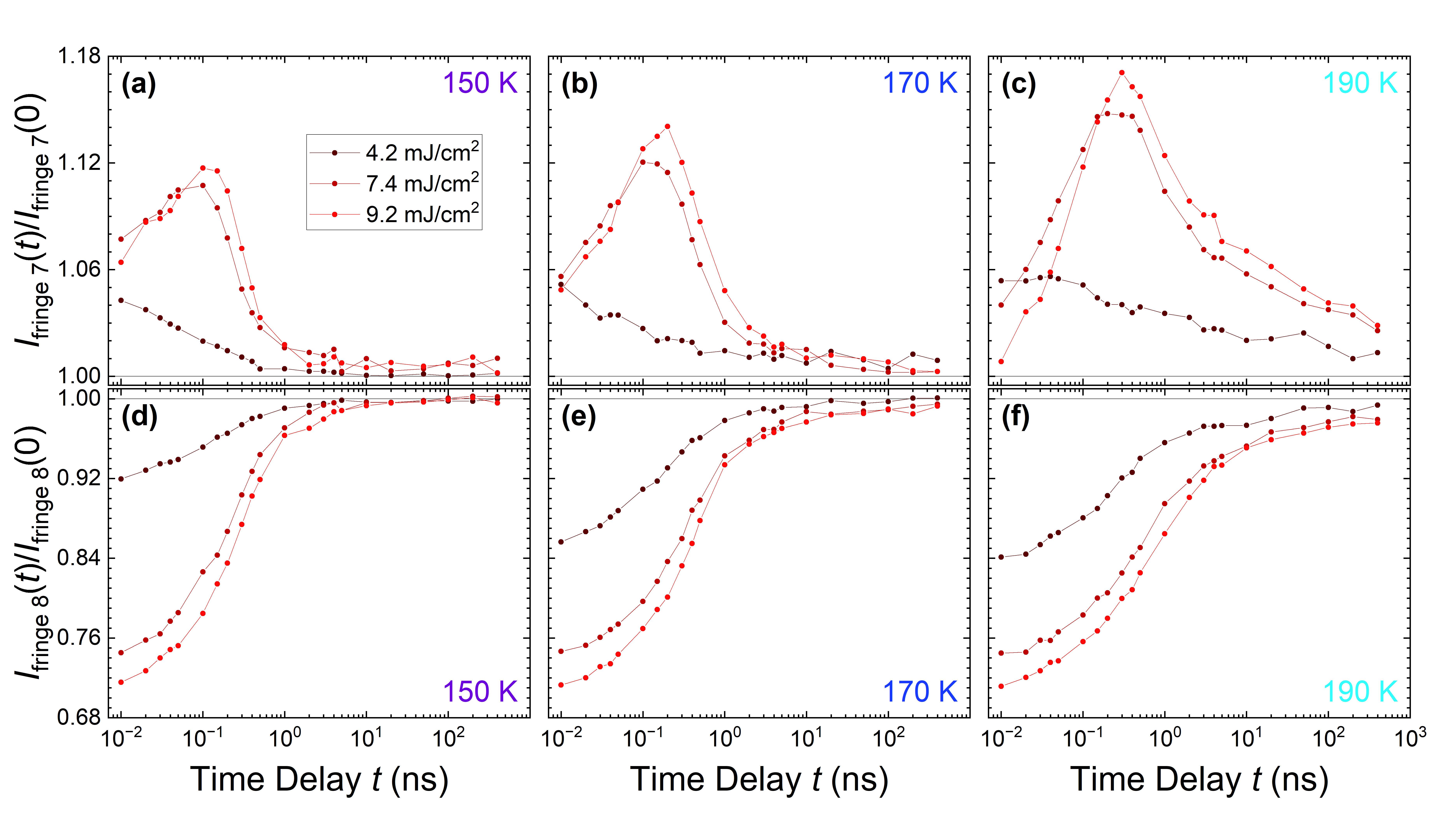}
    \caption{Normalized intensities of (a, b, c) fringe~7 and (d, e, f) fringe~8 following photoexcitation with three different laser fluences from (a, d) 150~K, (b, e) 170~K, and (c, f) 190~K. Fringe intensities are normalized to their initial intensity prior to photoexcitation.}
    \label{fig:FigS2}
\end{figure*}

Figure~\ref{fig:FigS2} compares the response of the \nth{7} and \nth{8}~fringes to different laser fluences at initial sample temperatures of 150~K, 170~K, and 190~K. The magnitude of the change of the fringes in each case is consistent with what would be expected from heating the sample by 75~K, 125~K, and 160~K for laser fluences of 4.2~mJ/cm$^2$, 7.4~mJ/cm$^2$, and 9.2~mJ/cm$^2$, respectively. In particular, the magnitude of the decrease in intensity of the \nth{8}~fringe with the initial laser pulse is comparable for the medium and high laser fluence, where it is expected that the satellite peak will completely disappear from the \nth{8}~fringe and appear on the \nth{7}~fringe. For the lowest fluence, the drop is much lower, as the sample is not heated to a temperature above the transition to 7.5~CDW wavelengths. At 190~K (see Fig.~\ref{fig:FigS2}(c)), the increase in intensity of the \nth{7}~fringe is initially lower for the highest fluence than for the medium and lowest fluence, consistent with heating the film above the Néel temperature where the satellite peak disappears. We note, however, that the time scales for recovery are independent of the laser fluence, changing only with the sample temperature. The recovery at all three fluences for an initial temperature of 150 K occurs on the order of 1~ns. This is increased by an order of magnitude at 170 K, and the reorientation has not occurred by 400 ns at any of the fluence at 190 K. Thus, the critical slowing depends on the initial sample temperature but not on the strength of the photoexcitation.

\begin{figure*}
    \centering
    \includegraphics[width=\textwidth]{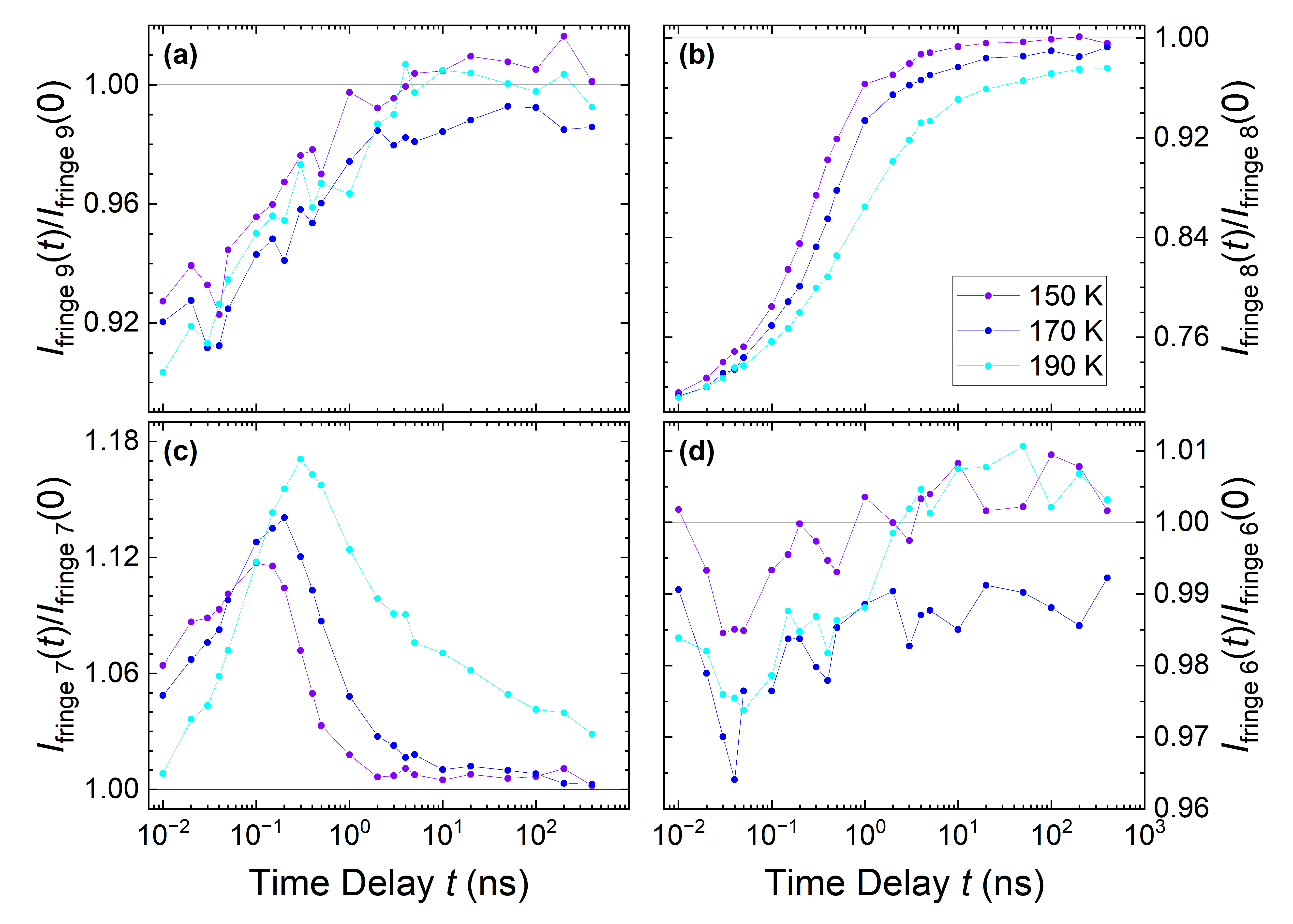}
    \caption{Normalized intensities of (a), fringe~9, (b) fringe~8, (c) fringe~7, and (d) fringe~6 following photoexcitation with fluence 9.2~mJ/cm$^2$ from different temperatures. Fringe intensities are normalized to their initial intensity prior to photoexcitation.}
    \label{fig:FigS3}
\end{figure*}

\begin{figure*}
    \centering
    \includegraphics[width=\textwidth]{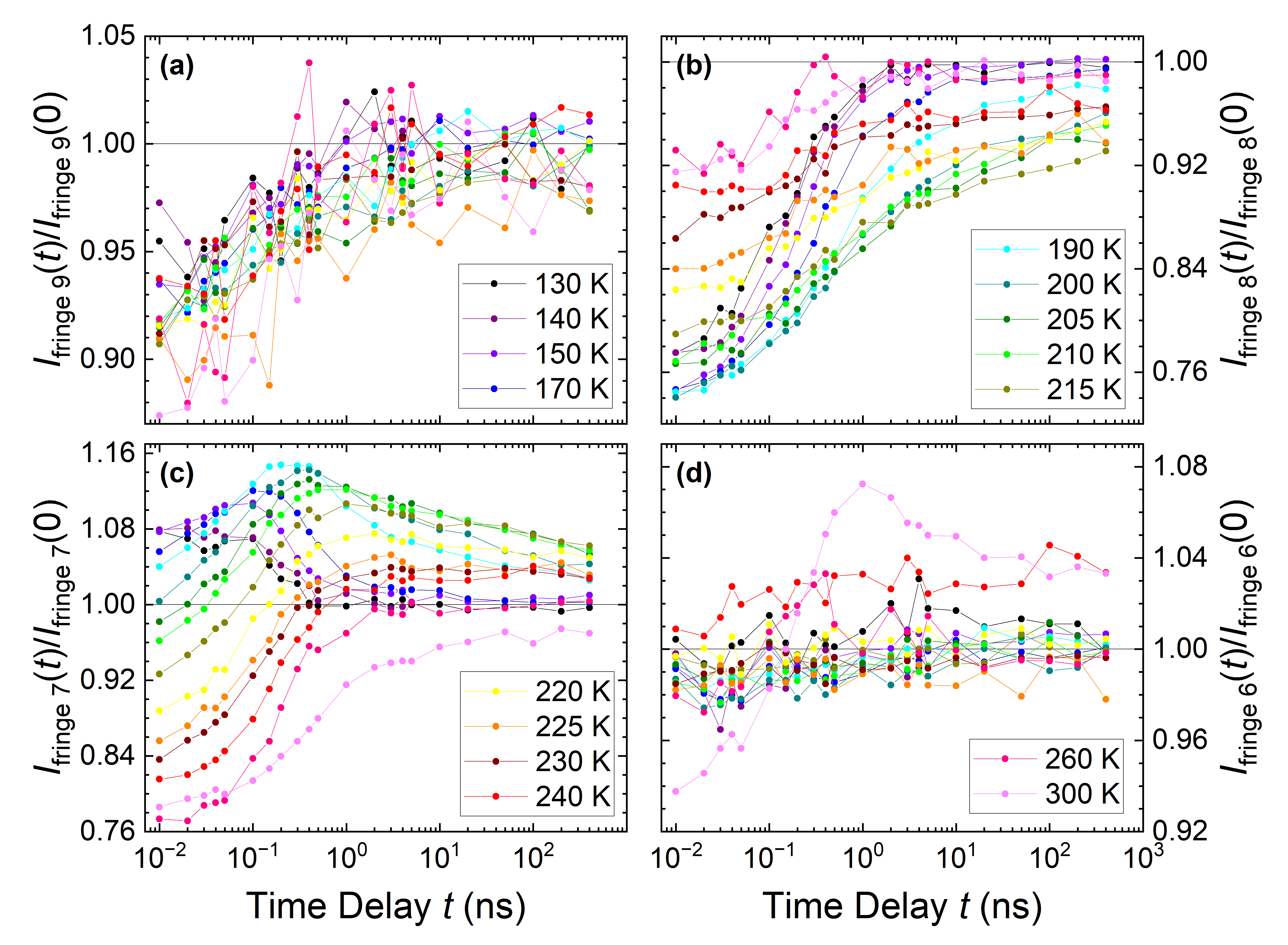}
    \caption{Normalized intensities of (a), fringe~9, (b) fringe~8, (c) fringe~7, and (d) fringe~6 following photoexcitation with fluence 7.4~mJ/cm$^2$ from different temperatures. Fringe intensities are normalized to their initial intensity prior to photoexcitation.}
    \label{fig:FigS4}
\end{figure*}

\begin{figure*}
    \centering
    \includegraphics[width=\textwidth]{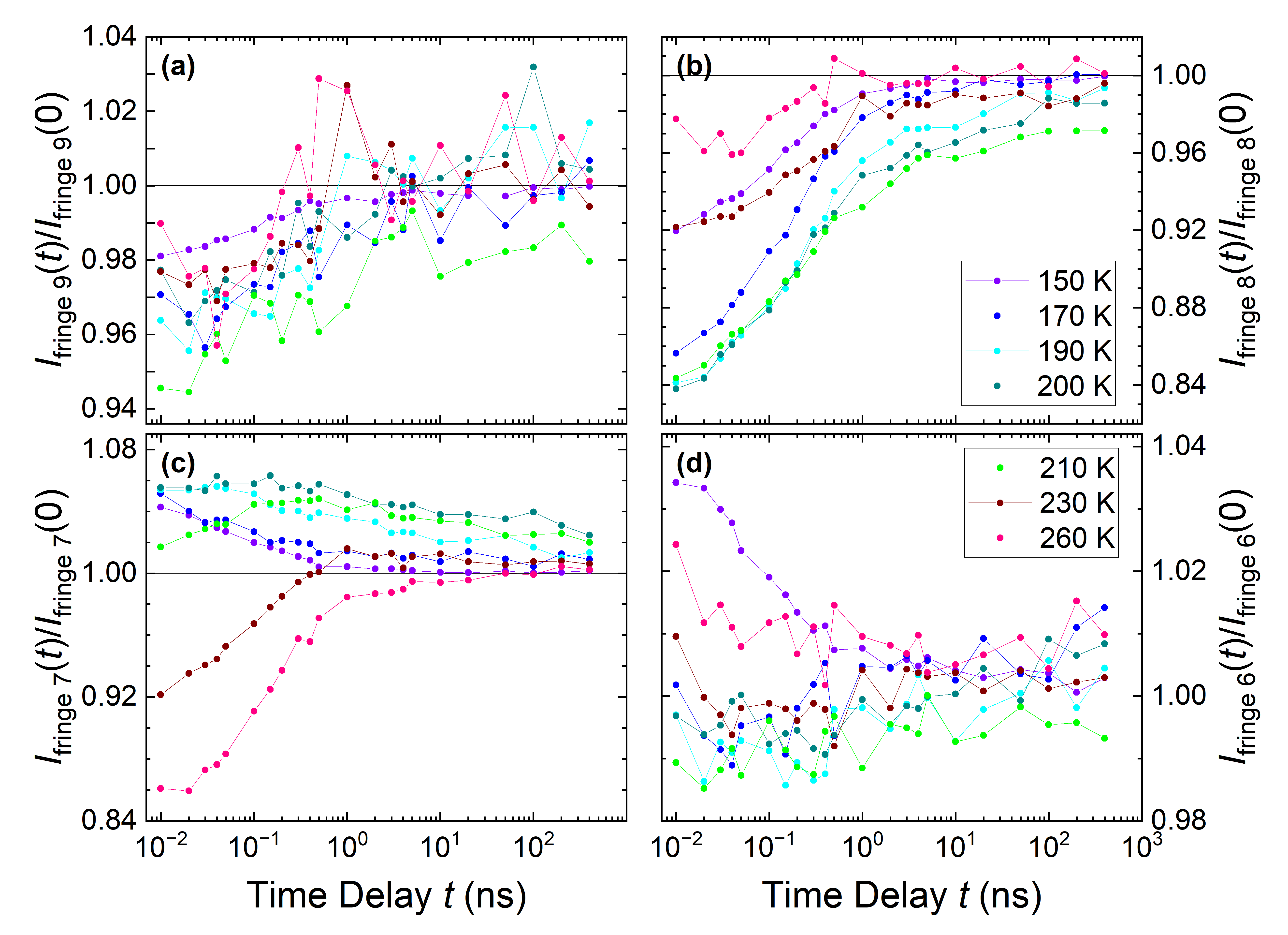}
    \caption{Normalized intensities of (a), fringe~9, (b) fringe~8, (c) fringe~7, and (d) fringe~6 following photoexcitation with fluence 4.2~mJ/cm$^2$ from different temperatures. Fringe intensities are normalized to their initial intensity prior to photoexcitation.}
    \label{fig:FigS5}
\end{figure*}

Figures~\ref{fig:FigS3}-\ref{fig:FigS5} show the complete set of responses of the \nth{9}, \nth{8}, \nth{7}, and \nth{6} fringes for the three laser fluences at all measured temperatures.

\section{Energy analysis of the data}

\begin{table}[!ht]

\caption[Parameters determined from fits of the recovery of the charge density wave amplitude following photoexcitation]{Parameters determined from fits of the recovery of the charge density wave amplitude following photoexcitation to Eq.~\ref{eq:Fits}. The initial sample temperature is $T$. Fitted fringe refers to which fringe amplitudes that was used for the fit. These amplitudes were the fringe intensities as a function of delay time normalized to their initial intensity prior to photoexcitation. $E_b$ is the energy barrier calculated from $\tau_\text{slow}$ assuming an attempt time of $\tau_0 = 250$~ps.}

\begin{center}
\begin{tabular}{c c c c c c c}%{p{1.5in} p{2.2in} p{2.2in}}

\toprule
\toprule

$T$ & Fitted & Fluence & $\tau_\text{fast}$ & $\tau_\text{slow}$ & \multirow{2}{*}{C} & $E_b$ \\

[K] & Fringe & [mJ/cm$^2$] & [ns] & [ns] & {} & [meV] \\

\midrule
\midrule

130 & 8 & 7.4 & $0.10 \pm 0.04$ & $0.43 \pm 0.17$ & $0.996 \pm 0.002$ & $6 \pm 5$ \\
140 & 8 & 7.4 & $0.17 \pm 0.04$ & $0.76 \pm 0.66$ & $0.992 \pm 0.004$ & $14 \pm 11$ \\
150 & 8 & 4.2 & $0.11 \pm 0.02$ & $0.80 \pm 0.21$ & $0.997 \pm 0.001$ & $15 \pm 4$ \\
150 & 8 & 7.4 & $0.23 \pm 0.02$ & $1.8 \pm 0.6$ & $0.999 \pm 0.002$ & $25 \pm 5$ \\
150 & 8 & 9.2 & $0.29 \pm 0.01$ & $3.7 \pm 0.8$ & $0.998 \pm 0.001$ & $35 \pm 3$ \\
170 & 8 & 4.2 & $0.17 \pm 0.02$ & $2.2 \pm 0.6$ & $0.997 \pm 0.001$ & $32 \pm 4$ \\
170 & 8 & 7.4 & $0.33 \pm 0.02$ & $3.8 \pm 0.7$ & $0.990 \pm 0.001$ & $40 \pm 3$ \\
170 & 8 & 9.2 & $0.43 \pm 0.02$ & $7.4 \pm 2.2$ & $0.988 \pm 0.002$ & $50 \pm 5$ \\
190 & 8 & 4.2 & $0.31 \pm 0.02$ & $17.1 \pm 6.7$ & $0.991 \pm 0.002$ & $69 \pm 7$ \\
190 & 8 & 7.4 & $0.48 \pm 0.04$ & $9.0 \pm 2.0$ & $0.977 \pm 0.002$ & $59 \pm 4$ \\
190 & 8 & 9.2 & $0.58 \pm 0.06$ & $10.5 \pm 2.7$ & $0.979 \pm 0.003$ & $61 \pm 5$ \\
200 & 8 & 4.2 & $0.28 \pm 0.03$ & $22.8 \pm 9.3$ & $0.985 \pm 0.003$ & $78 \pm 8$ \\
200 & 8 & 7.4 & $0.41 \pm 0.05$ & $9.5 \pm 2.2$ & $0.948 \pm 0.003$ & $63 \pm 5$ \\
205 & 8 & 7.4 & $0.44 \pm 0.06$ & $16.0 \pm 5.0$ & $0.937 \pm 0.004$ & $73 \pm 6$ \\
210 & 8 & 4.2 & $0.15 \pm 0.03$ & $2.7 \pm 0.5$ & $0.967 \pm 0.002$ & $43 \pm 4$ \\
210 & 8 & 7.4 & $0.38 \pm 0.04$ & $18.3 \pm 4.4$ & $0.945 \pm 0.003$ & $78 \pm 5$ \\
215 & 8 & 7.4 & $0.37 \pm 0.04$ & $20.5 \pm 6.4$ & $0.923 \pm 0.003$ & $82 \pm 7$ \\
220 & 8 & 7.4 & $0.18 \pm 0.05$ & $4.2 \pm 1.2$ & $0.941 \pm 0.003$ & $54 \pm 6$ \\
225 & 8 & 7.4 & $0.32 \pm 0.06$ & $58.0 \pm 54.8$ & $0.948 \pm 0.007$ & $106 \pm 19$ \\
230 & 8 & 4.2 & $0.34 \pm 0.06$ & $7.2 \pm 11.6$ & $0.990 \pm 0.002$ & $67 \pm 32$ \\
230 & 8 & 7.4 & $0.26 \pm 0.05$ & $9.7 \pm 7.2$ & $0.961 \pm 0.003$ & $73 \pm 15$ \\
\midrule
240 & 7 & 7.4 & $0.20 \pm 0.02$ & $2.1 \pm 1.1$ & $1.031 \pm 0.002$ & $44 \pm 11$ \\
260 & 7 & 4.2 & $0.17 \pm 0.02$ & $4.1 \pm 1.6$ & $1.000 \pm 0.002$ & $63 \pm 9$ \\
260 & 7 & 7.4 & $0.21 \pm 0.03$ & $2.2 \pm 1.9$ & $0.998 \pm 0.003$ & $48 \pm 20$ \\
300 & 7 & 7.4 & $0.47 \pm 0.03$ & $10.6 \pm 2.9$ & $0.969 \pm 0.002$ & $8 \pm 97$ \\

\bottomrule
\bottomrule
\end{tabular}
\end{center}
\label{tab:FitParam}
\end{table}

Figure S6 shows example fits of the data to the exponential decay function with two time constants:
\begin{equation}
\Delta\text{CDW} \sim A_1 \, \mathrm{e}^{-\frac{t}{\tau_\text{fast}}} + A_2 \, \mathrm{e}^{-\frac{t}{\tau_\text{slow}}} + C .
\label{eq:Fits}
\end{equation}
As the temperature approaches the hysteretic regions, the time scales slow and we are limited by the maximum achievable delay of 400 ns. The quality of the fits suffers as a result. Table~\ref{tab:FitParam} provides the parameters derived from fits to the data at all fluences and temperatures.

\begin{figure*}
    \centering
    \includegraphics[width=0.65\textwidth]{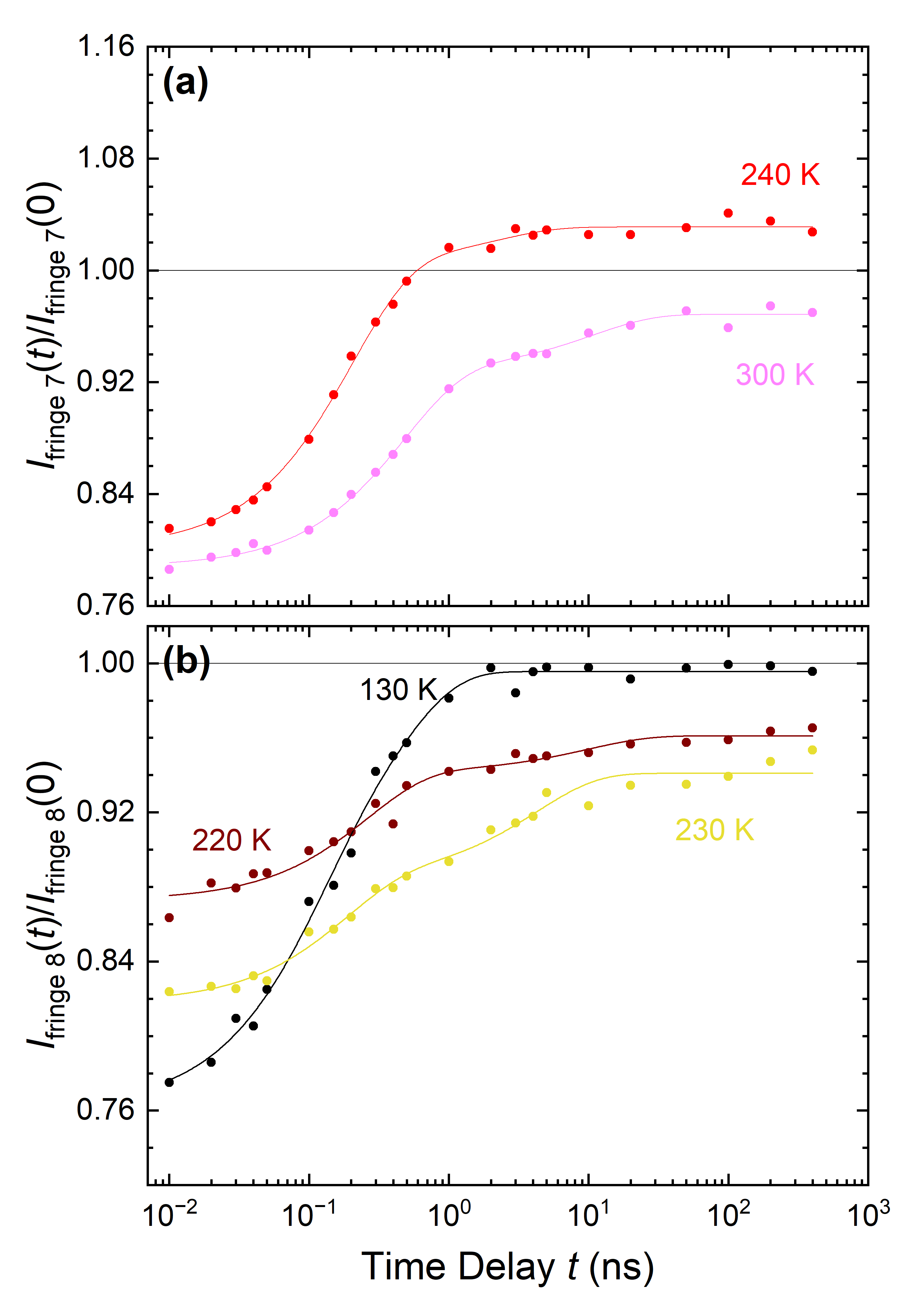}
    \caption{Some example normalized fringe intensities following photoexcitation along with fits to the data. See text for details of fit.}
    \label{fig:FigS6}
\end{figure*}

Energy barriers were calculated from the slow time constant of the fits with an Arrhenius model:
\begin{equation}
\ln\frac{\tau_\text{slow}}{\tau_0} = \frac{E_b}{k_B T} ,
\end{equation}
where $\tau_0$ is the attempt time. The attempt time for magnetic systems varies widely from $10^{-11}$ to 10$^{-8}$~s for different materials\footnote{T. Berndt, A. R. Muxworthy, and G. A. Paterson, “Determining the magnetic attempt time $\tau_0$, its temperature dependence, and the grain size distribution from magnetic viscosity measurements,” J. Geophys. Res.: Solid Earth, {\bf 120}, 7322-7336 (2015).} and is unknown for this system. Figure~S7 shows the calculated energy barrier for the reorientation from 7.5 to 8.5~CDW wavelengths for fluence 7.4~mJ/cm$^2$ for attempt times of 10~ps, 100~ps, 250~ps, and 1~ns. There is a small change in the temperature dependence associated with changing the attempt time, but the biggest difference we see in this range is a shift that changes the temperature at which the energy barrier goes to zero. For an attempt time of $\tau_0 = 1$~ns, a linear fit to the data shows the energy barrier goes to zero at about 135 K. For $\tau_0 = 250$~ps, the energy barrier would go to zero at 120~K, for $\tau_0 = 100$~ps at 100~K, and for $\tau_0 = 10$~ps at 80~K. We are unable to experimentally determine where the barrier goes to zero in this experiment because we cannot resolve a time constant smaller than the thermal time constant for recovery.

\begin{figure*}
    \centering
    \includegraphics[width=0.8\textwidth]{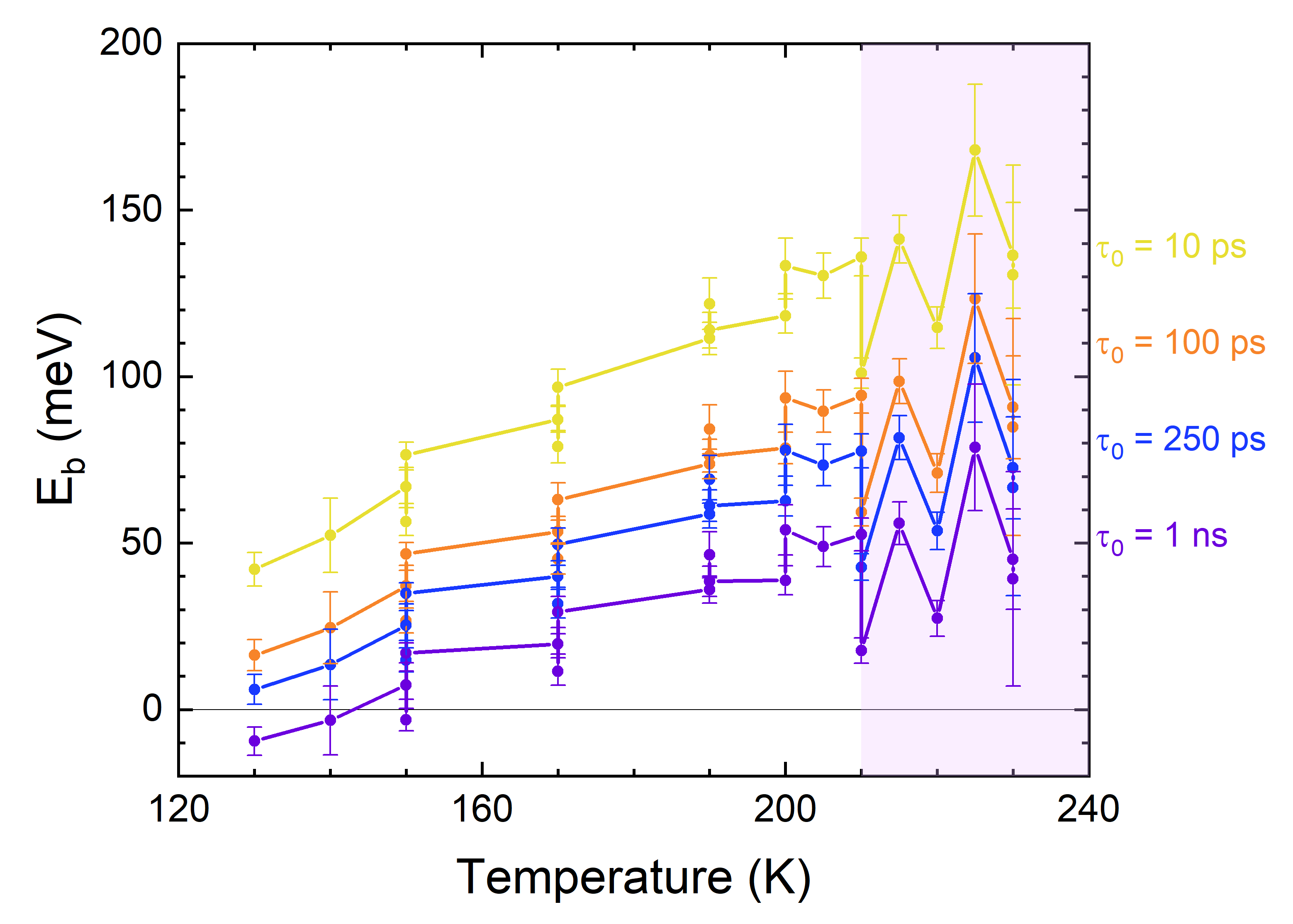}
    \caption{The calculated energy barrier for the reorientation from 7.5 to 8.5~CDW wavelengths for fluence 7.4~mJ/cm$^2$ for attempt times of 10~ps, 100~ps, 250~ps, and 1~ns.}
    \label{fig:FigS7}
\end{figure*}